# INSTRUMENTATION DEVELOPMENTS AND BEAM STUDIES FOR THE FERMILAB PROTON IMPROVEMENT PLAN LINAC UPGRADE AND NEW RFQ FRONT-END*

Victor E. Scarpine#, Cheng-Yang Tan, Pat R. Karns, Daniel S. Bollinger, Kevin L. Duel, Nathan Eddy, Ning Lui, Alexei Semenov, Raymond E. Tomlin, and William A. Pellico,
Fermilab, Batavia, IL 60510, USA

*Abstract*

Fermilab is developing a Proton Improvement Plan (PIP) to increase throughput of its proton source. The plan addresses hardware modifications to increase repetition rate and improve beam loss while ensuring viable operation of the proton source through 2025. The first phase of the PIP will enable the Fermilab proton source to deliver 1.8e17 protons per hour by mid-2013. As part of this initial upgrade, Fermilab plans to install a new front-end consisting of dual H- ion sources and a 201 MHz pulsed RFQ. This paper will present beam studies measurements of this new front-end and discuss new beam instrumentation upgrades for the Fermilab linac.

## INTRODUCTION

From its beginning, Fermilab has operated a successful program of supplying high-energy protons to its experimental programs. The present Fermilab proton facility (H- sources, pre-accelerator, linac and booster) has been operational for many years. However, this proton facility will be required to supply protons to the Main Injector and to the 8-GeV physics program until the era of Project X.

To meet these requirements, Fermilab has embarked on the Proton Improvement Plan (PIP) to improve and upgrade the present proton facilities [1]. The specific objectives are to enable proton operation capable of delivering 1.8E17 protons/hour (at 12 Hz) by 2013 and 2.25E17 protons/hour (at 15 Hz) by 2016 while maintaining Linac/Booster availability greater than 85%, maintaining residual activation at acceptable levels, and ensuring a useful operating life through 2025.

The first phase of the PIP is underway. Part of this first phase includes (1) replacing the present H- source and pre-accelerator with a new front-end injector based on a radio frequency quadrupole (RFQ) and (2) upgrading many of the linac beam diagnostics instrumentation. This work will be completed during the present long-term shutdown, which ends spring of 2013.

## NEW LINAC FRONT-END

The present linac front-end consists of dual Cockcroft-Walton pre-accelerators with individual H- ion sources. A 2009 review determined that this Cockcroft-Walton-based front-end is a liability and a large source of linac downtime. The first phase of the PIP project will replace the dual Cockcroft-Walton sources with an RFQ-based front-end. This new front-end injector consists of

- two 35 keV H- magnetron sources on a movable slide;
- a two-solenoid low-energy beam transport (LEBT) with an Einzel lens chopper with a beam current toroid between the solenoids;
- a 201.25 MHz, 750 keV RFQ; and
- a short medium-energy beam transport (MEBT) with a single buncher cavity.

This new front-end will feed beam into the first drift tube linac. Figure 1 shows a rendering of this new front-end injector, while figure 2 shows the actual ion source, LEBT and RFQ portion of the injector system.

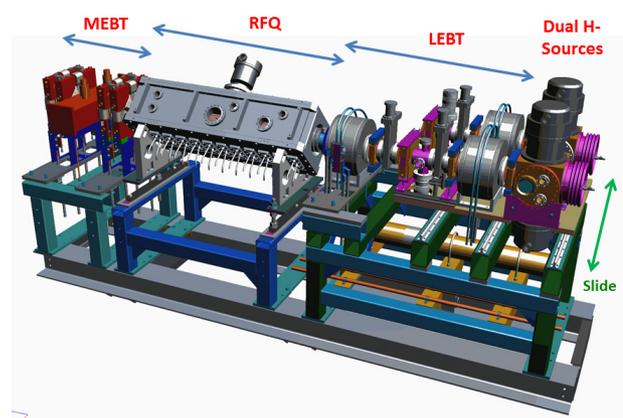

Figure 1: Fermilab's new front-end injector system.

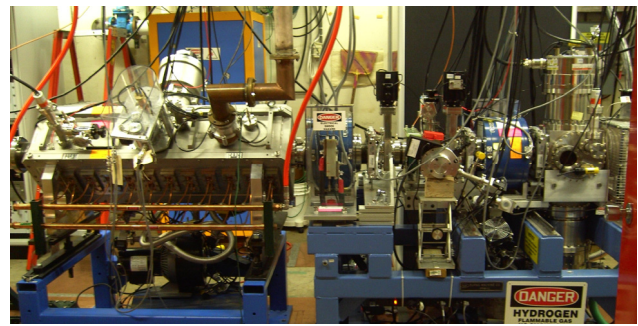

Figure 2: Completed ion source, LEBT and RFQ portion of the front-end injector system assembled in the source testing lab.

___________________________________________
*Work supported by U. S. Department of Energy under contract No. DE-AC02-07CH11359.
#scarpine@fnal.gov

## LEBT Commissioning Instrumentation

A number of beam measurements have been made in order to commission the ion source and LEBT portion of the injector system. These include
- beam current transmission and beam centroid steering through the two-solenoid LEBT, and
- beam size and transverse emittance near the entrance of the RFQ.

Figure 3 shows the various beam diagnostics setups used to make these measurements.

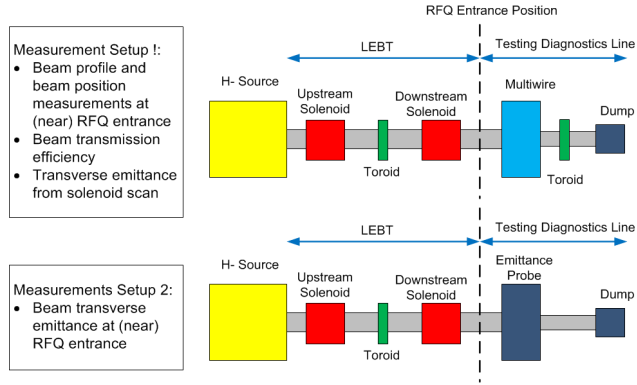

Figure 3: Schematic of measurements setups for commissioning through the LEBT

Two measurements of the LEBT performance include (1) a mapping of the beam transport through the LEBT to the entrance of the RFQ and (2) the transverse emittance at the entrance of the RFQ. Figure 4 shows a mapping of the beam centroid transverse position at the entrance of the RFQ as a function of the LEBT steering dipoles.

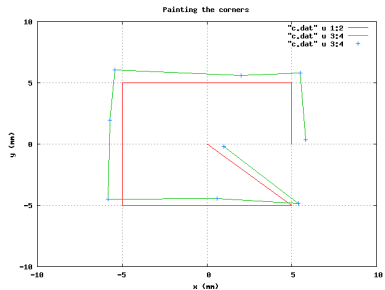

Figure 4: A comparison of the measurements range of the transverse beam position near the entrance of the RFQ (green) versus predicted (red) for different LEBT steering..

Figure 5 shows the horizontal and transverse phase space measurements of the LEBT beam near the entrance of the RFQ. A fit of these phase data yields a horizontal emittance of 0.21 pi-mm-mr and a vertical emittance of 0.17 pi-mm-mr. These values are one sigma normalized emittances.

The performance of the LEBT Einzel as a beam chopper is shown in figure 6. These measurements show a ~100 ns rise-time and fall-time for the chopped beam.

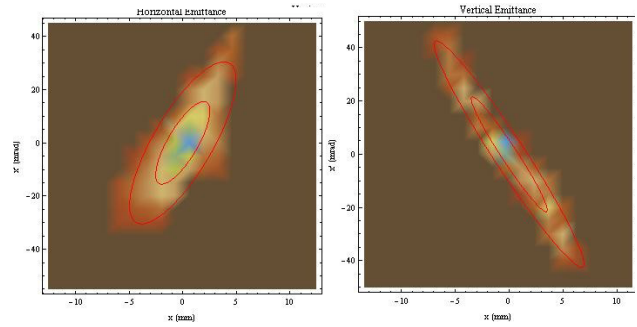

Figure 5: A measurement of the horizontal (left) and vertical (right) transverse phase space near the entrance of the RFQ. Red ellipses are one and two sigma fits.

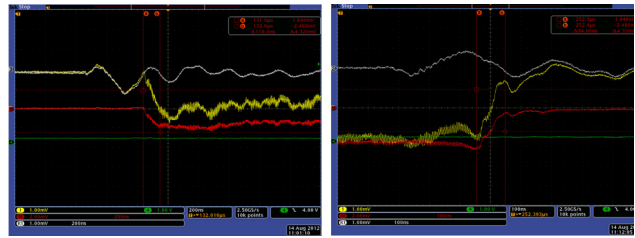

Figure 6: A beam toroid measurement of the leading-edge (left) and trailing-edge (right) of the Einzel lens beam chopping performance. The yellow curve is with-beam, the white curve is without-beam and the red curve is the difference.

## RFQ Commissioning Instrumentation

Table 1 lists the nominal design values and goals for the new Fermilab four-rod RFQ.

Table 1: RFQ Design Specifications

| Parameter | Value | Units |
|---|---|---|
| Input energy | 35 | keV |
| Output energy | 750 | keV |
| Frequency | 201.25 | MHz |
| Length | 102 | cm |
| Duty factor (80 μs, 15 Hz) | 0.12 | % |
| Design current | 60 | mA |

In order to measure the performance of the new RFQ, a number of beam measurements have been made under various operating conditions, e. g.. RFQ power and ion source settings. These RFQ performance measurements include
- beam current transmission;
- beam transverse profiles;
- transverse emittance;
- longitudinal bunching; and
- beam energy.

Figure 7 shows the various beams diagnostic setups used to make these measurements. This paper will present only the longitudinal bunching performance and the beam energy.

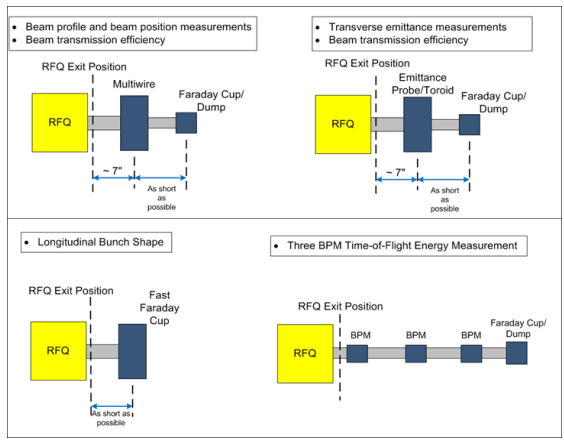

Figure 7: Schematic of measurements setups for commissioning through the RFQ.

## RFQ Bunching Performance

The bunching performance of the new RFQ is measured using a fast Faraday cup (FFC). The FFC is a broad bandwidth pickup capable of measurements up to 10 GHz. This bandwidth is adequate to qualify the bunching performance of the RFQ. Figure 8 shows the response of the FFC across the ~100 μs beam pulse for different values of RF power. This data shows that at low RF power, most of the beam is not bunched. Figure 9 shows histograms of the 1σ values of Gaussian fits to the individual bunches in figure 8. This shows that the 1σ longitudinal bunch size is ~420 ps for nominal RF power levels.

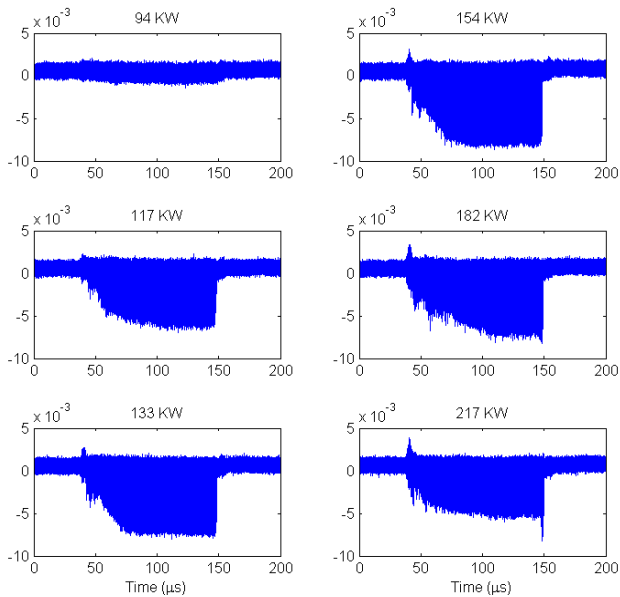

Figure 8: FFC signal across individual beam pulses as a function of RFQ RF power.

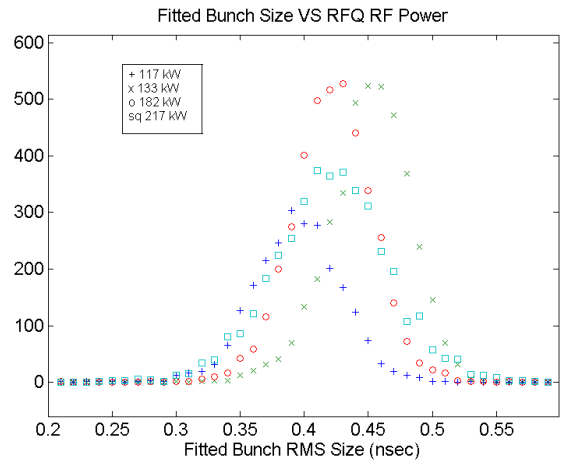

Figure 9: Histogram of 1σ values of Gaussian fits to individual bunches measured by the FFC for different RFQ RF power.

## RFQ Beam Energy

Initial RFQ beam energy was determined using time-of-flight (TOF) measurements utilizing a three-BPM system illustrated in figure 7. This TOF system was constructed with close-spacing and far-spacing BPMs to allow for coarse and fine energy measurements respectfully.

In general, TOF is a very precise method of measuring energy of non-relativistic beams. However, this precision relies on uniquely identify the same beam bunch at each BPM. The use of a three-BPM TOF system helps remove the need for unique bunch identification by relying on the same energy measurement for any pair of BPMs.

Figure 10 shows the results of TOF measurements for both 1st and 2nd harmonic signals from the BPMs. The figure plots the absolute energy, for each BPM pair combination, as a function the number of bunches between BPM pairs. The results give a best match beam energy of 705 keV. These measurements were also made over RFQ RF power levels with consistent results indicating that the RFQ beam energy is ~7% below the nominal design goal.

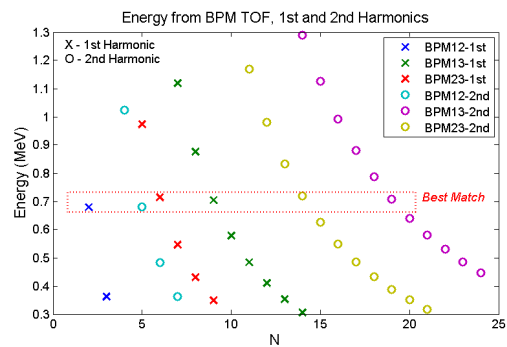

Figure 10: FFC signal across individual beam pulses as a function of RFQ RF power.

In order to verify the TOF energy measurements, a magnetic-based energy spectrometer was constructed. This spectrometer line includes two slits to collimate the beam, a dipole magnet to deflect the beam and a multi-wire to measure the deflection of the beam. Figure 11 shows the constructed energy spectrometer line.

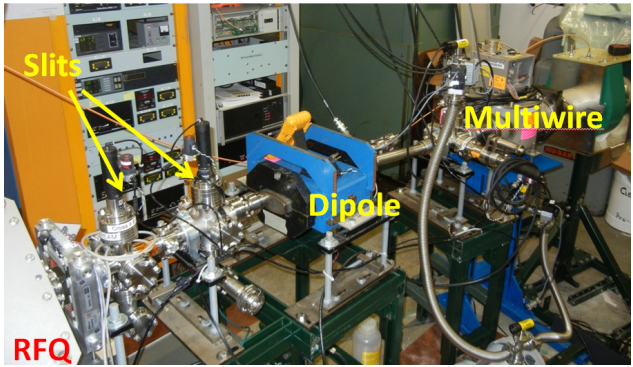

Figure 11: Energy spectrometer at end of RFQ.

A measurement of the RFQ beam energy with the spectrometer line yields a value of 701 keV and confirms the results of measurements by the TOF setup.

Additional studies of the RFQ design, along with beam and electromagnetic simulations, seem to indicate the cause of the low beam energy is due to a plate at the output end of the RFQ. It was determined that removal of this plate would still allow the RFQ to operate. Figure 12 shows the inside of the RFQ at the downstream end with and without this end plate. New energy spectrometer measurements, after removal of this end plate, now yield a beam energy of 756 keV, which is within the performance requirements for the RFQ.

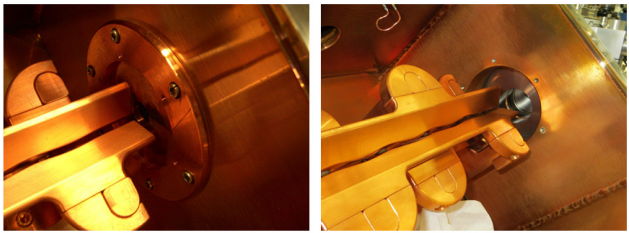

Figure 12: View of RFQ interior with (left) and without (right) end plate.

## LINAC BEAM DIAGNOSTIC UPGRADES

In addition to replacing the injector front-end, Fermilab will upgrade the electronics for the linac BPM and toroid systems.

### Linac BPM System Upgrade

The existing linac BPM RF electronics will be replaced with a digital electronics readout system. This is being done to add long-term stability and future flexibility to the BPM system. For this upgrade, no changes will be made either to the BPM detectors or to the signal cables.

Fermilab has been developing custom FPGA-based digital electronics for a number of accelerator applications [2]. Following these earlier applications, these new linac BPM electronics include high-speed ADCs and digital signal processing via a programmable FPGA.

The new electronics will provide average beam position, intensity and relative phase over each beam pulse for every BPM at a rate of 15 Hz. This information will be acquired by the Fermilab ACNET via the linac controls system. The on-board FPGA digital processing will implement phase measurements with a resolution of 0.1 degrees at 201 MHz. This phase information can then be used for time-of-flight energy measurements aong the linac. Table 2 lists linac beam parameters and BPM measurement requirements. In addition, these new electronics will implement a calibration system to improve long-term stability.

Table 2: Linac BPM Specifications

| Parameters | Min. | Nominal | Max. |
|---|---|---|---|
| Beam intensity | 5 mA | 34 mA | 60mA |
| Beam frequency | | 201.24 MHz | |
| Position meas. range | | +/- 50 mm | |
| Beam rise time | | | 200 ns |
| Beam pulse duration | 2.2 μs | 25 μs | 100 μs |
| Position resolution | | 0.1 mm | |
| Phase resolution | | 0.1 degrees | |
| Long-term position stability | | 0.25 mm | |

### Linac Toroid System Upgrade

Similar to the upgrade of the linac BPM system, Fermilab will replace the present toroid electronics with a digital electronics readout system [3]. This gives the similar benefits of long-term stability and future flexibility to the toroid system as well. These new toroid electronics consist of a custom designed VME digitizer board with eight signal channels each utilizing a 14 bit, 125 mega-samples per second ADC. This board also includes a FPGA to provide such digital signal processing as pulse integration, baseline correction and automatic pulse edge detection. Figure 13 shows a block diagram of this new VME digitizer board.

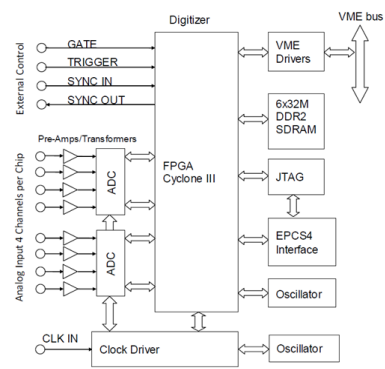

Figure 13: Block diagram of the new 8-channel digitizer VME board.

## ACKNOWLEDGMENT

We would like to acknowledge the following people for their contributions to this work. From Fermilab, B. Schupbach, K. Koch, A. Feld, P. Balakrishnan (summer student, MIT), B. Ogert, J. Briney, J. Kubinski, K. Triplett, J. Larson, R. Mraz,, J. Lackey, G. Velev, A. Makarov, V. Kashikhin, J. DiMarco and G. Romanov, B. Oshinowo, C. Wilson, G. Coppola, G. Teafoe and the entire survey group. We would also like to thank J. Schmidt, B. Koubek, A. Schempp of the University of Frankfurt and S. Kurennoy of LANL.